\documentclass[11pt,a4paper]{article}
\usepackage{jcappub}

\usepackage{latexsym}
\usepackage{epsfig}
\usepackage{amssymb}
\usepackage{amsmath}
\usepackage{dcolumn}
\usepackage{bm}
\usepackage{graphicx}
\usepackage{subfig}

\usepackage{amsfonts,ulem,wrapfig,epsfig}

\newcommand{\lp}{\left(}
\newcommand{\rp}{\right)}
\newcommand{\lb}{\left[}
\newcommand{\rb}{\right]}

\newcommand{\ba}{\begin{eqnarray}}
\newcommand{\ea}{\end{eqnarray}}
\newcommand{\be}{\begin{equation}}
\newcommand{\ee}{\end{equation}}

\newcommand{\al}{\alpha}

\begin{document}

\title{The Cosmology of Interacting Spin-2 Fields}

\author{Nicola Tamanini}
\affiliation{Department of Mathematics,
University College London,
Gower Street, London, WC1E 6BT, UK}
 \author{Emmanuel N. Saridakis}
 \affiliation{Physics Division, National Technical University of Athens,
15780 Zografou Campus,  Athens, Greece}
\affiliation{Instituto de F\'{\i}sica, Pontificia Universidad de Cat\'olica
de Valpara\'{\i}so, Casilla 4950, Valpara\'{\i}so, Chile}
\author{Tomi S. Koivisto}
\affiliation{Institute for Theoretical Astrophysics, University of Oslo,
N-0315 Oslo, Norway}
 
\date{\today}

\abstract{
We investigate the cosmology of interacting spin-2 particles, formulating
the multi-gravitational theory in terms of vierbeins and without imposing
any Deser-van Nieuwen-huizen-like constraint. The resulting
multi-vierbein theory represents a wider class of gravitational theories if
compared to the corresponding multi-metric models. Moreover, as opposed to
its metric counterpart which in general seems to contain ghosts, it has
already been proved to be ghost-free. We outline a discussion about the
possible matter couplings and we focus on the study of cosmological scenarios
in the case
of three and four interacting vierbeins. We find rich behavior, including de
Sitter solutions 
with an effective cosmological constant arising from the multi-vierbein 
interaction,
dark-energy solutions and nonsingular bouncing behavior.}

\maketitle

\section{Introduction}

The formulation of a consistent theory of interacting spin-2 field, as a
well-defined theoretical problem, has a long history \cite{Isham:1971gm}.
These constructions in general suffer from the same Boulware-Deser ghost
instability \cite{Boulware:1973my} that plagues non-linear extensions of
massive gravity, since one can show that not only they allow but they indeed
demand for at least one massive spin-2 field \cite{Boulanger:2000rq}.
Recently  a consistent way to suitably choose the
(self)interacting terms in order to raise the cutoff of the effective theory
and systematically remove the  Boulware-Deser (BD) ghost  for the case of one
massive spin-2 field   was found \cite{deRham:2010ik,deRham:2010kj}.
Therefore,  one could follow the same procedure in order to cure the BD
instabilities for the case of two-gravity theories too
\cite{Hassan:2011tf,Hassan:2011zd,Hassan:2011ea}.

The advantage of these non-linearly completed ``bi-metric'' theories is
that they allow for consistent cosmological solutions in agreement with
observations
\cite{Volkov:2011an,vonStrauss:2011mq,Volkov:2012cf,Berg:2012kn,
Volkov:2012zb,Akrami:2013pna,
Nojiri:2012zu,Capozziello:2012re,Mohseni:2012ug,Nojiri:2012re,Maeda:2013bha,
Volkov:2013roa},
while their single-metric, nonlinear
massive-gravity counterparts, although safe at the fundamental level, have
been found to exhibit instabilities at the cosmological perturbation level
\cite{DeFelice:2012mx,D'Amico:2011jj,
DeFelice:2013awa,DeFelice:2013bxa,Gumrukcuoglu:2013nza} \footnote{Although
there is still a discussion whether bi-metric theories exhibit similar
problems with nonlinear massive gravity, their different field  content, as
well as their distinct dynamics and predictions, does not allow for the
immediate transfer of the arguments between the two cases
\cite{Kuhnel:2012gh,Akrami:2012vf,Tasinato:2012ze,Deser:2012qx,Kluson:2013cy,
Deser:2013uy,deRham:2013wv,Hassan:2013pca,Deser:2013gpa,Deser:2013eua,
Kluson:2013aca}.}. Thus, it is
very interesting to examine the cosmological applications of gravitational
theories where more than two
interacting gravitational sectors appear \cite{Hinterbichler:2012cn}.

A crucial comment must be made at this point. Up to now, the majority of the
above works in multi-gravitational theories have been performed in the
metric-language, giving rise to bi-metric, tri-metric constructions etc.
Although one can formulate them using the vierbeins as the fundamental fields
instead, it was recently realized that the opposite is not always true,
namely that the multi-vierbein constructions can be more general, without a
corresponding equivalent metric formulation. For instance this is the case of
the ``triangle'' interaction of three gravitational sectors, which although
safe at the vierbein level it  potentially contains ghost when formulated in
terms of metrics \cite{Hinterbichler:2012cn}.

More generally the equivalence between metric and vielbein formulations
relies on a particular condition (sometime called the Deser-van Nieuwenhuizen
gauge \cite{Deser:1974cy} for bi-metric theories) which has to be imposed on 
the vielbeins
\cite{Hassan:2012wt}. Although in the case of massive gravity
\cite{Ondo:2013wka}, or at the perturbative level of multi-metric gravity
\cite{Hinterbichler:2012cn}, this condition arises from the field equations,
in the general case of more than two interacting gravitational sectors it is
not clear whether this is still true
\cite{Deffayet:2012zc,Alexandrov:2012yv}, or if it has to be imposed  as
a separate
assumption.

Hence, if the Deser-van Nieuwenhuizen  condition is not assumed then 
the vielbein formulation of the theory corresponds
to a different physical theory with respect to the common ghost-free
bi-metric
gravity. In other words, starting from the vielbein formulation of
\cite{Hinterbichler:2012cn} without imposing any condition on the tetrads,
one ends up with a theory which has no known corresponding metric formulation,
and thus is in principle physically different. The un-restricted vielbein
approach seems thus to describe a much
wider class of theories which can be used to characterize interacting
gravitational sectors.
Such qualitative advantages of the vierbein
formulation should be taken into account in the constructions of
gravitational theories, and moreover they may enlighten the discussion of
which field is the fundamental one, especially proceeding towards the
quantization of the theory\footnote{New ways of defining the gravitational
degrees of freedom have been explored recently, for instance in the Cartanian
framework 
of \cite{Westman:2012xk,Westman:2012zk}, and in the contexts  of the bimetric
variational principle \cite{Koivisto:2011vq,BeltranJimenez:2012sz} and doubly
connected spacetimes \cite{Tamanini:2012mi,Koivisto:2013kwa}.}.
  
In the present work we are interested in investigating the cosmology of
interacting spin-2 fields. The use of the multi-vierbein formulation allows
us to incorporate interaction terms that were missed in earlier multi-metric
works \cite{Khosravi:2011zi,Nomura:2012xr,Zinoviev:2013hac}.
We find that even considering only extra terms which have no known
metric description, it can lead to a very rich cosmology.   
  The plan of the paper is as follows:  In section \ref{Basicmodel} we
present
the multi-vierbein gravitational theories and in section \ref{mattercoupling}
we discuss the incorporation of the matter sector.
In section \ref{model} we focus on the cosmology of three interacting
spin-2 fields, extracting analytical solutions at both inflationary as well
as late-times eras, while in section \ref{fourvierbein} we perform the
analysis for four interacting gravitons. Finally, section
\ref{Conclusions} is devoted to the discussion and summary of the obtained
results. 

{\it Notation}. Greek letters are used for world (manifold) indices, while $a,b,c,...$ indices are used for local (tangent space) indices. Both take values from 0 to 3 and are summed when repeated. Indices running from $i,j,k,...$ are used to number the $N$ tetrads in the theory and are not summed if repeated unless explicitly specified. A tetrad field is related to the corresponding metric through
\begin{align}
g_{\mu\nu} = \eta_{ab}\,e^a_\mu\,e^b_\nu \,,
\end{align}
and we use the convention $\eta_{\mu\nu}=\eta_{ab}={\rm diag}(-,+,+,+)$.

\section{Multi-vielbein Action and Field Equations}
\label{Basicmodel}

We now briefly present the formulation of multi-gravitational theories
describing $N$
interacting spin-2 fields in 4 dimensions \cite{Hinterbichler:2012cn}. The
corresponding action reads 
\begin{align}
S = \int \sum_i \mathcal{L}^{(i)}_{\rm EH}  + \epsilon_{abcd} \sum_{i,j,k,l}
\tilde\chi^{ijkl}\, {\bf e}_{(i)}^a \wedge {\bf e}_{(j)}^b \wedge {\bf
e}_{(k)}^c \wedge {\bf e}_{(l)}^d \,,
\label{001}
\end{align}
where the Latin indices $i,j,k,l$ run from 1 to $N$ and $\tilde\chi^{ijkl}$
is a completely symmetric tensor of constant coefficients. The   first term
provides the 
Einstein-Hilbert Lagrangian for every tetrad, given by
\begin{align}
\label{tetrad_eh}
\mathcal{L}^{(i)}_{\rm EH}=\frac{M_i^2}{4} {\bf e}_{(i)}^a \wedge {\bf
e}_{(i)}^b \wedge \star {\bf R}_{ab}^{(i)} = \frac{M_i^2}{2} \det e_{(i)}\,
R^{(i)} \,,
\end{align}
where $M_i$ represents the Planck mass, ${\bf R}_{ab}^{(i)}$ the Ricci tensor
2-form and $R^{(i)}$ the scalar curvature of the $i$th tetrad field, while
$\star$ denotes the Hodge dual. The second term in (\ref{001}) accounts for
the interaction terms that at most contain four tetrads.

In principle we could additionally consider parity-odd
interacting terms such
as, for example,
\begin{align}
e_{(i)}^a \wedge e_{(i)}^b \wedge e_{(j)}{}_a \wedge e_{(j)}{}_b \,.
\end{align}
However, since the ghost-freedom has been proven only  for action (\ref{001})
\cite{Hinterbichler:2012cn} and in order to keep the analysis as simple as
possible, without loss of generality
we will neglect parity-odd interactions in the present work.  In any case,
for a first cosmological
application the parity-even terms of (\ref{001}) are adequate to capture all
the novel features of the theory\footnote{This is because the isotropic and 
homogeneous background would not
allow nontrivial contributions from the parity-odd terms.  In passing we note
however that the possible role of those
terms could be interesting to explore in view of the parity-odd anomalies
observed in the cosmic microwave background \cite{Ade:2013nlj}.  Previously
they have been attempted to be generated by extended gravity by assuming
(metric) Chern-Simons modifications \cite{Alexander:2006mt} or noncommutativity 
of space-time \cite{Koivisto:2010fk}.}.

In \cite{Hinterbichler:2012cn,Deffayet:2012zc} it has been proven that for
$N=2$ (two gravitational sectors) and imposing the Deser-van Nieuwenhuizen
condition
\begin{align}
e_{(1)}^a{}_{[\mu}e_{(2)}{}_{\nu]a} = 0 \,,
\label{099}
\end{align}
the theory (\ref{001}) is equivalent to the ghost-free bi-metric gravity more
commonly considered in the metric formulation
\cite{Hassan:2011zd,Hassan:2011ea}. If this condition is not imposed, then
the equivalence of the two-metric with the two-vielbein formulation is not
guaranteed. On the other hand, in \cite{Alexandrov:2012yv} it was shown that
this condition could be restricting, since it leads to  two massless
propagating gravitons as in two exact copies of General Relativity. As we
mentioned in the Introduction, although in the case of massive
gravity \cite{Ondo:2013wka}, or at the perturbative level of multi-metric
gravity \cite{Hinterbichler:2012cn}, this condition arises from the field
equations themselves, in the general non-perturbative case of more than two
interacting gravitational sectors it is not clear whether this is still true
\cite{Deffayet:2012zc,Alexandrov:2012yv}, or if it has to be imposed  as
a separate assumption. Therefore, in the present work we prefer not to impose
the Deser-van Nieuwenhuizen condition or any other constraint on the tetrad
fields. Clearly, this leads to a much wider class of multi-gravitational
theories, and to formulations of multi-vierbein theories with no known
corresponding metric formulation, even for only two interacting
vielbeins. The resulting multi-vierbein theory is not a mere re-formulation of
the multi-metric constructions, but a new, richer multi-gravitational
theory.

Varying action (\ref{001}) with respect to the tetrad one-form ${\bf
e}_{(i)}^a$ produces the field equations
\begin{align}
\frac{M_i^2}{2} {\bf e}_{(i)}^b \wedge \star {\bf R}_{ab}^{(i)} +
\epsilon_{abcd} \sum_{j,k,l} \chi^{ijkl}\, {\bf e}_{(j)}^b \wedge {\bf
e}_{(k)}^c \wedge {\bf e}_{(l)}^d =0 \,,
\label{003}
\end{align}
which can be rewritten as
\begin{eqnarray}
&&M_i^2 G_{(i)}{}^\mu_\nu=- e_{(i)\nu}^a
e_{(i)e}^\mu\epsilon_{abcd}\,\epsilon^{efgh}    \sum_{i,j,k} \chi^{ijkl}
X_{(ij)}{}_f{}^b X_{(ik)}{}_g{}^c X_{(il)}{}_h{}^d,
\label{004}
\end{eqnarray}
where $G_{(i)}{}^\mu_\nu = R_{(i)}{}_\nu^\mu-1/2\delta^\mu_\nu R_{(i)}$ is
the usual Einstein tensor of the $i$th tetrad,   $\chi^{ijkl}$ is a new
completely symmetric tensor of coefficients directly related to
$\tilde\chi^{ijkl}$, and we have defined the matrices
\begin{align}
X_{(ij)}{}_a{}^b = e_{(i)}^\mu{}_a e_{(j)}^b{}_\mu =
(e_{(j)}e_{(i)}^{-1}){}_a{}^b \,.
\end{align}
The details of the derivation of the field equations can be found in Appendix
\ref{app:A}.



\section{Matter Coupling}
\label{mattercoupling}

In a multi-metric gravitational theory the issue arises to which metric
matter
fields should couple. The viable forms of the interactions between  
metrics are symmetric with respect to replacing one metric with another, 
and in principle one could consider matter fields coupling also symmetrically
to all the metrics \cite{Hassan:2011zd,Khosravi:2011zi}, as it was
recently realized in the bi-metric gravity 
\cite{Akrami:2013ffa} \footnote{Theories including two
separate matter sectors, each coupling 
exclusively to one of the metrics, have been also proposed. However, they
lead
inevitably to problems, such as violations of the energy conditions  and
inherent degeneracy in interpreting the observables
\cite{Baccetti:2012re,Capozziello:2012re}.}. We mention here that going
beyond the simple one-metric-matter coupling leads to violation of the
equivalence principle, and thus such generalized couplings are potentially strongly
constrained by experiments, and in particular from those related to the
equivalence principle violation. Due to the Vainshtein screening mechanism
at play in these gravity theories, it is not yet however clear how strong the 
constraints will be. 

However, apart from this experimental
requirement, at the theoretical level such couplings are allowed, and thus
it would be interesting to consider them too in the following discussion,
along with the simple one-metric-matter coupling,
as a first step towards understanding their implications. 
When considering couplings that are linear in the matter lagrangians just as in
GR, it appears that by construction the system is devoid of ghosts. Technically
this is due to the matter lagrangian being linear in the lapse functions of each
respective metric, and thus preserving the crucial constraint nature of the 
equations of motions for the lapse functions. This constraint kills one progagating
degree of freedom, that would otherwise become the notorious Boulware-Deser ghost.
Considering more generic possible forms of the matter coupling however would
introduce nonlinear dependence on the lapse functions, resulting in the loss of the
constraint and thus allowing the ghost to propagate.
Generalised couplings to matter was briefly discussed in \cite{Hassan:2011zd}, 
and further in \cite{Hassan:2012wr} the possibility of coupling matter to the massless
combination of the metrics was explored. Since then matter lagrangian is not linear
in the lapse functions of the metrics, this choice for the coupling did not turn out
to be viable, as expected.

In this work we will generalize
the framework of multiply-coupled matter to theories with more than two
gravitational sectors, and moreover, having in mind the above discussion, we
will formulate it in terms of the vierbeins. As we will see, this can be
done only when the gravitational fields are non minimally coupled, that is 
$N$ copies  of non-interacting Einstein-Hilbert theories would be either
physically equivalent to General Relativity or inconsistent. 
 
Let us begin by extending  the action (\ref{001}) as 
\begin{eqnarray}
&&S = \int \sum_i \mathcal{L}^{(i)}_{\rm EH}  + \epsilon_{abcd}
\sum_{i,j,k,l}
\tilde\chi^{ijkl}\, {\bf e}_{(i)}^a \wedge {\bf e}_{(j)}^b \wedge {\bf
e}_{(k)}^c \wedge {\bf e}_{(l)}^d  +  \int \sum_i c_i
\mathcal{L}^{(i)}_{m}(\Phi,{\bf e}_{(i)}) \,,
\label{actionfull}
\end{eqnarray}
where the sums run from $i=1$ to $i=N$ with $N$ the number of the
different vierbeins in
the theory. In this case, the terms
\begin{align}
S^{(i)}_{\rm matter} = \int c_i
\mathcal{L}^{(i)}_{m}(\Phi,{\bf e}_{(i)}) \,,
\end{align}
account for the matter action, with
$\Phi$ denoting the matter fields collectively, and the dimensionless
coupling constants $c_i$ the relative strength of the coupling of each tetrad
to matter. 
Thus, we consider the total action to include $N$ copies of the matter
Lagrangian, each having the same functional form. This is $N$-times minimally
coupled theory in the sense that each $i$-term separately would reduce to the
standard theory.

Varying the action (\ref{actionfull}) with respect to each of the tetrads we
obtain
$N$ field equations as
\begin{align}
\frac{M_i^2}{2} {\bf e}_{(i)}^b \wedge \star {\bf R}_{ab}^{(i)} +
\epsilon_{abcd} \sum_{j,k,l} \chi^{ijkl}\, {\bf e}_{(j)}^b \wedge {\bf
e}_{(k)}^c \wedge {\bf e}_{(l)}^d =c_i T_a^{(i)} \,,
\end{align}
where the energy-momentum 3-form coupled to the $i$th tetrad is defined as
\begin{align}
T_a^{(i)} = \frac{\delta S_{\rm matter}}{\delta e^a_{(i)}}  =\det
e_{(i)}T_a^{(i)\mu} \epsilon_{\mu\nu\alpha\beta}\,dx^\nu\wedge
dx^\alpha\wedge dx^\beta \,.
\end{align}
These field equations can be rewritten more conveniently as
\begin{eqnarray}
&&M_i^2 G_{(i)}{}^\mu_\nu + e_{(i)\nu}^a
e_{(i)e}^\mu\epsilon_{abcd}\,\epsilon^{efgh}   \sum_{j,k,l} \chi^{ijkl}
X_{(ij)}{}_f{}^b X_{(ik)}{}_g{}^c X_{(il)}{}_h{}^d = c_i
T^{(i)\mu}_{\phantom{(i)}\nu},\ \
\label{fieldeqnmatter}
\end{eqnarray}
which generalize (\ref{004}).
In this expression, $T^{(i)}_{\mu\nu}$ is the stress energy tensor for the
$i$th tetrad defined by
\begin{equation} 
T^{(i)\nu}_\mu = T^{(i)a}_\mu\,e_{(i)a}^\nu \,.
\end{equation}
If the matter sector can be formulated in terms of metrics this
energy-momentum tensor corresponds to the standard one defined in metric
General Relativity, namely
\begin{align}
T^{(i)}_{\mu\nu}=\frac{-2}{\sqrt{-g_{(i)}}}\frac{\delta \lp
\sqrt{-g_{(i)}}\mathcal{L}_m^{(i)}\rp}{\delta g_{(i)}^{\mu\nu}}\,.
\label{set}
\end{align}
On the other hand the definition with tetrads is more general since it
allows to incorporate fields such as spinors that cannot be coupled to gravity
as naturally, or indeed at all, within the metric formulation.

Next, we rewrite equation (\ref{fieldeqnmatter}) as
\begin{equation} 
M_i^2 G_{\mu\nu}^{(i)} +W_{\mu\nu}^{(i)}  = {c_i}T^{(i)}_{\mu\nu} \,,
\end{equation}
where the short-hand $W_{\mu\nu}^{(i)}$ denotes the interaction terms with
the other vierbeins in the field equations. At the level of field equations
the
new consequence of multiple matter coupling is, as expected, that the matter
fields
now appear as sources in all the equations.

Additionally, note that the multiple coupling affects the behavior of
matter fields, too. In order to see this more transparently we consider
the conservation laws. Since each Einstein tensor is covariantly
conserved with respect to its
covariant derivative, it follows that
\be \label{cons}
\nabla^\mu_{(i)}  W_{\mu\nu}^{(i)} = \nabla^\mu_{(i)} {c_i}T^{(i)}_{\mu\nu}
\,.
\ee
On the other hand, as the total matter sector should be diffeomorphism
invariant, it obeys, as a whole, the conservation law
\be \label{div}
\sum_i \det e_{(i)} \nabla^\mu_{(i)} {c_i}T^{(i)}_{\mu\nu} = 0\,.
\ee
Therefore, from the   two separate conservation equations
(\ref{cons}),(\ref{div}),    we obtain the constraint  
\be \label{cons_sum}
\sum_i \frac{1}{\det e_{(i)}}\nabla^\mu_{(i)} W^{(i)}_{\mu\nu} = 0\,.
\ee
Hence, despite the fact that the $W$-tensors are now not separately
conserved, the system is consistent, and   the field equations just 
imply   that
the sum (\ref{cons_sum}) vanishes. 
Note that in the case where the matter fields are coupled to only one tetrad,
this becomes a stronger set of $N$ constraints, $\nabla^\mu_{(i)}
W^{(i)}_{\mu\nu} = 0$, and then clearly particles would follow the
geodesics of
the one tetrad/metric they are coupled to. On the other hand, in the case of $N$
pure Einstein-Hilbert theories, that is when one   sets the $W$-tensors to
zero, equation (\ref{cons}) would force all matter to follow the geodesics of
all the metrics simultaneously. In that case the metrics/tetrads would coincide
with each other\footnote{Perhaps some solutions would exist in the case
where the metrics were the same only
up to an affine transformation that leaves the geodesics invariant.}, and
effectively the theory would reduce to General Relativity. 

In general, expression (\ref{cons}) seems to imply that matter does not
follow the geodesics of any tetrad. From the viewpoint of any given metric/tetrad,
the multiply-coupled theory predicts violations of the equivalence principle.
One expects that this fact could be used to impose strong constraints on the
coupling constants $c_i$. The equations of motion  for matter fields must
be consistent with the conservation law (\ref{cons}) and they will thus
involve couplings in principle to all tetrads $i$ for which $c_i \neq 0$. 

As
a concrete example, as a matter Lagrangian we consider the Lagrangian of a
canonical scalar field
$\phi$: 
\be \label{scal}
\mathcal{L}_\phi \equiv \sum_i c_i \mathcal{L}_\phi^{(i)}=\sum_i c_i \lb
\frac{1}{2}g_{(i)}^{\mu\nu}\phi_{,\mu}\phi_{ ,\nu} - V(\phi) \rb\,,
\ee 
where the $i$th metric is obviously given by
\begin{align}
g^{\mu\nu}_{(i)} = \eta^{ab}\,e_{(i)a}^\mu\,e_{(i)b}^\nu \,.
\end{align}
The equations of motion can be derived by the usual Euler-Lagrange method by
varying the action with respect to the scalar field\footnote{One should be
careful to vary the full action and not the Lagrangians, since now the
different measures of the metrics
give different weights to the contributions of the corresponding metrics.},
and
we find
\be \label{kg}
\sum_i c_i \det e_{(i)} \lb \Box^{(i)}\phi - V'(\phi)\rb = 0\,,
\ee
where $\Box^{(i)}$ is the dAlembertian operator of the metric $g_{(i)}$. On
the other hand, inserting the Lagrangian (\ref{scal}) into the stress-energy
tensor (\ref{set}) and then into the conservation laws
(\ref{cons}),(\ref{div}),(\ref{cons_sum}),  we obtain
\begin{equation}
\sum_i \det e_{(i)} \nabla^\mu_{(i)} {c_i}T^{(i)}_{\mu\nu} =  
\sum_i c_i \det e_{(i)} \lb \Box^{(i)}\phi - V'(\phi)\rb \phi_{,\nu} = 0\,.
\end{equation}
Thus, the conservation laws are   guaranteed to hold due to the
generalized Klein-Gordon equation
(\ref{kg}) for the scalar field and do not introduce any new constraint. In
summary, we have thus seen how the multiple coupling consistently modifies
both the structure of the gravitational equations and the equations of motion
for the matter fields. 

Finally, we mention that  the case of two metrics coupled
to dust-like
matter in cosmological spacetimes was investigated in detail in
\cite{Akrami:2013ffa}, and it was
found that each term in the sums (\ref{div}) and (\ref{cons_sum}) vanishes
independently. Thus, it remains to be seen how this features generally occurs, and how stringent are the constraints
ensuing from the violations of the usual conservation
laws when
it does not occur.

\section{Three-Vierbein Cosmology}
\label{model}

\begin{figure}[ht]
\subfloat[][Full interacting
theory]{\includegraphics[width=0.45\columnwidth]{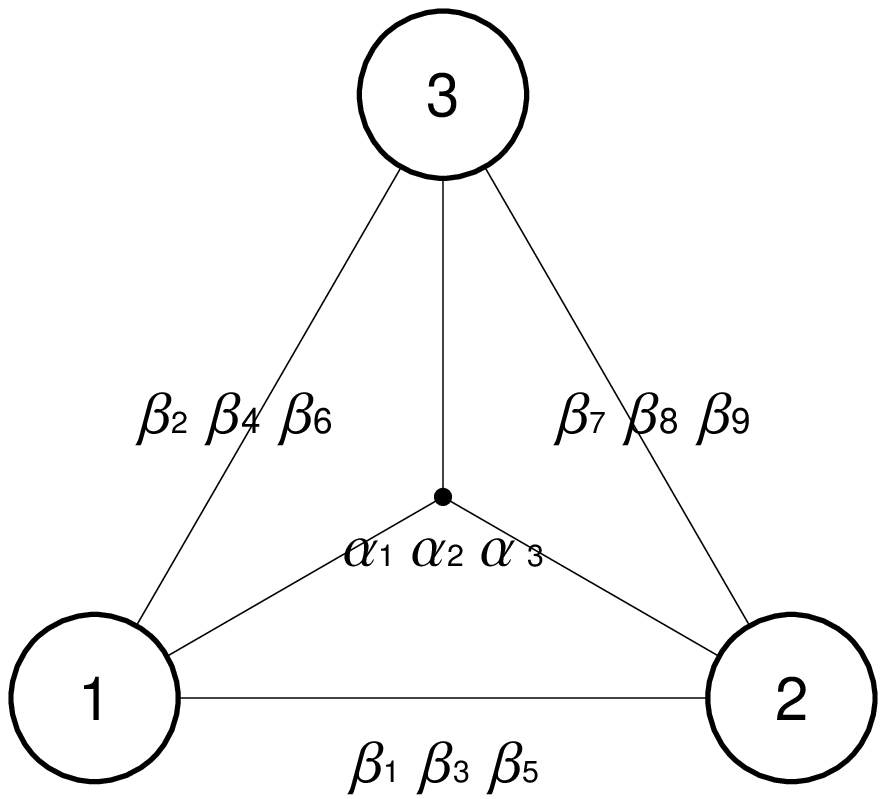}} \qquad
\subfloat[][Three-vertex interacting
theory]{\includegraphics[width=0.45\columnwidth]{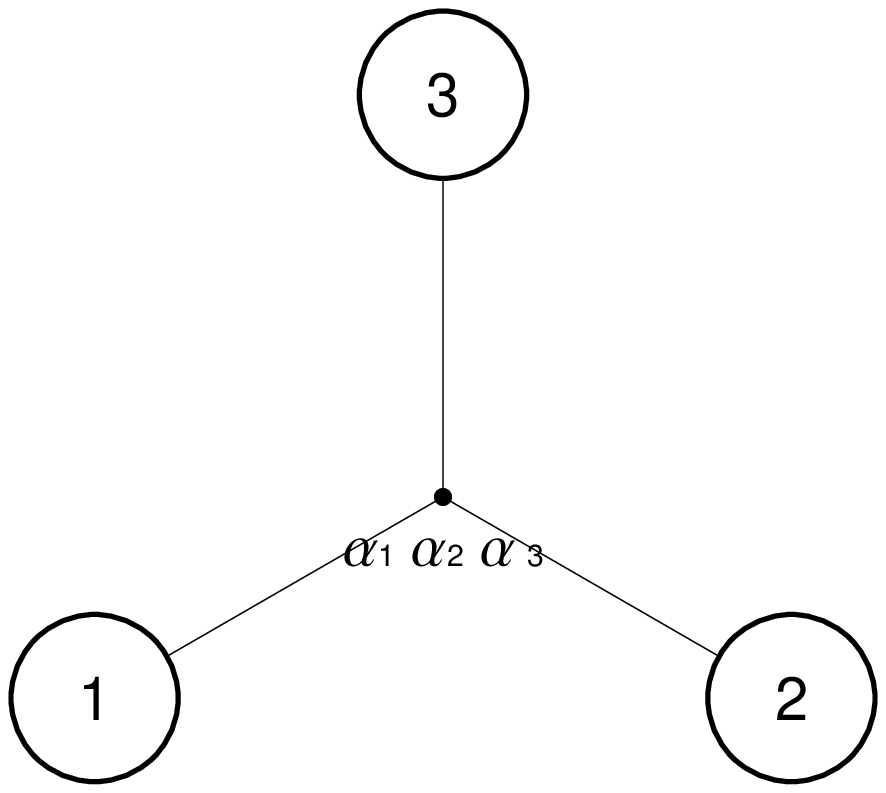}}
\caption{
Graphic representation of the three-vierbein theory in four
dimensions.  The $\beta$-edges represent the terms coupling pair of tetrads,
while the $\alpha$ three-vertex junctions represent the terms coupling three
tetrads together. The $\Lambda$-terms (cosmological constants)  
correspond to self-interacting/closed lines and have not be drawn.
}
\label{fig:1}
\end{figure}

Let us now investigate the cosmological applications of the multi-vierbein
gravity formulated above. In this section we focus on three interacting
tetrads, while in the next one we will study the four interacting vierbein
scenario. Assuming three interacting gravitational sectors, that is setting 
 $N=3$ in the expressions of the previous section,  we result in  15 coupling
terms in four dimensions, namely: 3 cosmological constant terms ($\Lambda_1$,
$\Lambda_2$, $\Lambda_2$), 9 terms coupling pairs of tetrads ($\beta_1$,
...~,~$\beta_9$) and 3 terms coupling
all the three tetrads together ($\alpha_1$, $\alpha_2$, $\alpha_3$). The
theory can be graphically visualized as in Fig.~\ref{fig:1}~(a), where the
coupling parameters are associated to the corresponding edge.

In order to investigate the cosmological evolution in a universe governed by
the three-interacting spin-2
field theory, we assume that the three tetrads take
the form 
\begin{eqnarray} 
&&e_{(1)}{}_\mu^a=\mbox{diag}\left(N_a(t), \frac{a(t)}{\sqrt{1-k\,r^2}},
a(t)\,
r, a(t)\,r\,\sin\theta\right) \label{008}\nonumber \\
&&e_{(2)}{}_\mu^a=\mbox{diag}\left(N_b(t), \frac{b(t)}{\sqrt{1-k\,r^2}},
b(t)\,
r, b(t)\,r\,\sin\theta\right) \label{009}\nonumber \\
&&e_{(3)}{}_\mu^a=\mbox{diag}\left(N_c(t), \frac{c(t)}{\sqrt{1-k\,r^2}},
c(t)\,
r, c(t)\,r\,\sin\theta\right),\ \ \ \ \ \label{010}
\end{eqnarray}
corresponding to the three Friedmann-Robertson-Walker (FRW) metrics
\begin{eqnarray}
&&ds^2_1 = -N_a(t)^2\,dt^2 + a(t)^2 \left(\frac{dr^2}{1-k r^2}+r^2
d\Omega^2\right)
 \label{005}\nonumber \\
&&ds^2_2 = -N_b(t)^2\,dt^2 + b(t)^2 \left(\frac{dr^2}{1-k r^2}+r^2
d\Omega^2\right)
 \label{006}\nonumber \\
&&ds^2_3 = -N_c(t)^2\,dt^2 + c(t)^2\left(\frac{dr^2}{1-k r^2}+r^2
d\Omega^2\right)
 ,\label{007}
\end{eqnarray}
where $N_a$, $N_b$, $N_c$, $a$, $b$, $c$ (the three
lapse functions and the three scale factors) are all functions of $t$,
$k=0,\pm 1$ as usual and
$d\Omega^2=d\theta^2+\sin^2\theta\,d\phi^2$ is the 2-sphere line element. 
Note that, similarly to the bi-metric gravity case, we are considering that
the spatial curvature of all the three
metrics is the same. If this assumption is relaxed then inconsistencies arise
from the field equations, 
similarly to the bi-metric
gravity case \cite{DeFelice:2012mx,D'Amico:2011jj,
DeFelice:2013awa,DeFelice:2013bxa,Gumrukcuoglu:2013nza}. Furthermore,
although as usual we could eliminate one lapse function redefining the time
$t$, for clarity and generality we   keep all the $N_i$ in the forthcoming
expressions, having in mind that  one of them can be set to unity at any
moment.

We mention here that  the FRW assumptions for the tetrads indeed solve the
Deser-van Nieuwenhizen condition (\ref{099}). This means that even though
the theory we are dealing with does not contain such condition in general, at
the cosmological level such restriction is automatically satisfied. As a
consequence, all the results we will find in this and the following sections
will hold also in the corresponding metric description of the theory.

Finally, a comment should be made  concerning the vierbein
choice corresponding
to a specific metric. As it is known, due to the local Lorentz invariance
there are infinite vierbein choices producing a given metric. Since in the
multi-vierbein gravity of action (\ref{001}) the interacting terms do not
involve derivatives of the tetrads,  one can straightforwardly verify that it
is local Lorentz invariant and thus any Lorentz-transformed, non-diagonal, vierbein
choice would lead to the same field equations. Therefore, in this work we
consider diagonal vierbeins for simplicity and without loss of generality.
However, note that this is not {\it{a priori}} possible in any modified
gravitational theory, since in the case where local Lorentz invariance is broken
the vierbeins should be chosen very carefully (for instance in the case of
$f(T)$ gravity \cite{Tamanini:2012hg,Tamanini:2013xya}).


\subsection{Vacuum solutions}
\label{vacuumsolthree}

Let us first investigate the vacuum solutions of the theory, that is we
consider the field equations (\ref{004}) without the matter sector of
section \ref{mattercoupling}. Inserting the tetrads (\ref{010}) inside the
field equations (\ref{004}) yields the
following three first Friedmann equations
\begin{eqnarray}
&&3M_1^2\left(\frac{k}{a^2}+\frac{H_a^2}{N_a^2}\right)= \Lambda_1 +
3\beta_1\frac{b}{a} + 3\beta_2\frac{c}{a} +3\beta_3\frac{b^2}{a^2}\nonumber\\
&&\ \ \ +3\beta_4\frac{c^2}{a^2} + \beta_5\frac{b^3}{a^3}
+\beta_6\frac{c^3}{a^3} +
2\alpha_1\frac{b c}{a^2} +\alpha_2\frac{b^2 c}{a^3}+ \alpha_3\frac{b
c^2}{a^3}  \label{011},\nonumber\\
&&3M_2^2\left(\frac{k}{b^2}+\frac{H_b^2}{N_b^2}\right)= \Lambda_2
+\beta_1\frac{a^3}{b^3} +3\beta_3\frac{a^2}{b^2} +3\beta_5\frac{a}{b}
\nonumber\\
&&\ \ \ +3\beta_7\frac{c}{b} +3\beta_8\frac{c^2}{b^2} +\beta_9\frac{c^3}{b^3}
+\alpha_1\frac{a^2c}{b^3} +2\alpha_2\frac{ac}{b^2} +\alpha_3\frac{ac^2}{b^3}
 \label{012},\nonumber\\
&&3M_3^2\left(\frac{k}{c^2}+\frac{H_c^2}{N_c^2}\right)= \Lambda_3
+\beta_2\frac{a^3}{c^3} +3\beta_4\frac{a^2}{c^2} +3\beta_6\frac{3a}{c}\
\nonumber\\
&&\ +\beta_7\frac{b^3}{c^3} +3\beta_8\frac{b^2}{c^2}+3\beta_9\frac{b}{c}
\alpha_1\frac{a^2b}{c^3} +\alpha_2\frac{ab^2}{c^3} +2\alpha_3\frac{ab}{c^2}
,\ \ \label{013}  
\end{eqnarray}
plus the three acceleration equations
\begin{eqnarray}
&&
\frac{M_1^2}{N_a^2}
\left(N_a^2\frac{k}{a^2}+2\dot{H}_a+3H_a^2-2H_a\frac{\dot{N}_a}{N_a}\right)
 =\ \ \ \ \nonumber\\
&&\ \  \Lambda_1+ \beta_1\left(2\frac{b}{a}+\frac{N_b}{N_a}\right)
+\beta_2\left(2\frac{c}{a}+\frac{ N_c}{N_a}\right)
 \nonumber\\
&&
\ \ +\beta_3\left(\frac{b^2}{a^2}+2\frac{bN_b}{aN_a}\right)
+\beta_4\left(\frac{c^2}{a^2}+2\frac{cN_c}{aN_a}\right)  \nonumber\\
&& \ \
+\beta_5\frac{b^2N_b}{a^2N_a} +\beta_6\frac{c^2N_c}{a^2N_a}\nonumber\\
&& \ \  
+\frac{2\alpha_1}{3}\left(\frac{bc}{a^2}
+\frac{bN_c}{aN_a}+\frac{cN_b}{aN_a}\right)
+\frac{\alpha_2}{3}\left(\frac{b^2N_c}{a^2N_a}+2\frac{bcN_b}{a^2N_a}\right)
\nonumber\\  
&& \ \
+\frac{\alpha_3}{3}\left(\frac{c^2N_b}{a^2N_a}+2\frac{bcN_c}{a^2N_a}\right),\
\
 \label{014}
\end{eqnarray}
\begin{eqnarray}
&& \frac{M_2^2}{N_b^2}
\left(N_b^2\frac{k}{b^2}+2\dot{H}_b+3H_b^2-2H_b\frac{\dot{N}_b}{N_b}\right)
 =\ \ \ \ \nonumber\\
&& \ \   \Lambda_2 +\beta_1\frac{a^2N_a}{b^2N_b}
+\beta_3\left(\frac{a^2}{b^2}+2\frac{aN_a}{bN_b}\right)
\nonumber\\
&& \ \   +\beta_5\left(2\frac{a}{b}+\frac{N_a}{N_b}\right)
+\beta_7\left(2\frac{c}{b}+\frac{N_c}{N_b}\right)
\nonumber\\
&& \ \   
+\beta_8\left(\frac{c^2}{b^2}+2\frac{cN_c}{bN_b}\right)
+\beta_9\frac{c^2N_c}{b^2N_b}
\nonumber\\
&&  \ \  
+\frac{\alpha_1}{3}\left(\frac{a^2N_c}{b^2N_b}+2\frac{acN_a}{b^2N_b}
\right)
+\frac{2\alpha_2}{3}\left(\frac{ac}{b^2}+\frac{aN_c}{bN_b}
+\frac{cN_a}{bN_b}\right)\nonumber\\
&&  \ \  
+\frac{\alpha_3}{3}\left(\frac{c^2N_a}{b^2N_b}+2\frac{acN_c}{b^2N_b}\right),
 \label{015}  
\end{eqnarray}
\begin{eqnarray}
&&\frac{M_3^2}{N_c^2}
\left(N_c^2\frac{k}{c^2}+2\dot{H}_c+3H_c^2-2H_c\frac{\dot{N}_c}{N_c}\right)
 =\ \ \ \ \nonumber\\
&&  \ \  
\Lambda_3 +\beta_2\frac{a^2N_a}{c^2N_c}
+\beta_4\left(\frac{a^2}{c^2}+2\frac{aN_a}{cN_c}\right)
+\beta_6\left(2\frac{a}{c}+\frac{N_a}{N_c}\right)
 \nonumber\\ 
&&  \ \  +\beta_7\frac{b^2N_b}{c^2N_c}
+\beta_8\left(\frac{b^2}{c^2}+2\frac{bN_b}{cN_c}\right)
+\beta_9\left(2\frac{b}{c}+\frac{N_b}{N_c}\right)
\nonumber\\  
&&  \ \ 
+\frac{\alpha_1}{3}\left(\frac{a^2N_b}{c^2N_c}+2\frac{abN_a}{c^2N_c}\right)
+\frac{\alpha_2}{3}\left(\frac{b^2N_a}{c^2N_c}+2\frac{abN_b}{c^2N_c}\right)
 \nonumber\\ 
&&  \ \ 
+\frac{2\alpha_3}{3}\left(\frac{ab}{c^2}+\frac{aN_b}{cN_c}
+\frac{bN_a}{cN_c}\right) \,,\label{016}
\end{eqnarray}
where $H_a=\dot{a}/a$, $H_b=\dot{b}/b$, $H_c=\dot{c}/c$ are the three Hubble
functions and an overdot
denotes differentiation with respect to $t$.  The parameters appearing in
(\ref{011})-(\ref{016}) are related to the completely symmetric tensor of
coefficients $\chi^{ijkl}$ of equation (\ref{003})  through 
\begin{align}
&\Lambda_1=6\,\chi^{1111}\,, \quad \Lambda_2=6\,\chi^{2222}\,, \quad
\Lambda_3=6\,\chi^{3333}\,,  
\nonumber\\
&\beta_1=6\,\chi^{1112}\,, \quad \beta_2=6\,\chi^{1113}\,, \quad
\beta_3=6\,\chi^{1122}\,, 
\nonumber\\
& \beta_4=6\,\chi^{1133}\,, \quad
\beta_5=6\,\chi^{1222}\,, \quad  \beta_6=6\,\chi^{1333}\,,   
\nonumber\\
&
 \beta_7=6\,\chi^{2223}\,, \quad
\beta_8=6\,\chi^{2233}\,, \quad
\beta_9=6\,\chi^{2333}\,,  
\nonumber\\
&
\alpha_1=18\,\chi^{1123}\,, \quad \alpha_2=18\,\chi^{1223}\,, \quad
\alpha_3=18\,\chi^{1233}\,. \label{019}
\end{align}
From these expressions we deduce that the $\Lambda_i$ are just
the cosmological constants for the three vierbeins and do not couple
different fields, the $\beta$-terms couple
pair of fields, and finally the $\alpha$-terms acount for the ``triangular''
interactions of all the three vierbeins, as was conveniently visualized   in
Fig.~\ref{fig:1}~(a).

As we discussed in the Introduction,  the $\alpha$-terms, which are the fully
interacting ones, do not have a known metric corresponding description,
especially if we do not apply the  Deser-van Nieuwenhuizen condition or any
other constraint on the tetrad fields.
That is why although the simple interacting $\beta$ and
$\Lambda$ terms were considered in \cite{Khosravi:2011zi} within a metric
formulation of the theory, the full interaction was necessarily neglected. 

In this work we are interested in exactly these full interacting terms, and
in the novel features they bring to cosmology. Thus, we focus on a theory
where only the $\alpha$-terms are non vanishing, that is to the
model depicted in Fig.~\ref{fig:1}~(b). Obviously, one can straightforwardly
study the full theory too.

With these assumptions the Friedmann
equations (\ref{013}) read
\begin{eqnarray}
&&3M_1^2\left(\frac{k}{a^2}+\frac{H_a^2}{N_a^2}\right)= 2\alpha_1\frac{b
c}{a^2} +\alpha_2\frac{b^2 c}{a^3}+ \alpha_3\frac{b c^2}{a^3} 
,\nonumber\\
&&3M_2^2\left(\frac{k}{b^2}+\frac{H_b^2}{N_b^2}\right)=
\alpha_1\frac{a^2c}{b^3} +2\alpha_2\frac{ac}{b^2} +\alpha_3\frac{ac^2}{b^3}
,\nonumber\\
&&3M_3^2\left(\frac{k}{c^2}+\frac{H_c^2}{N_c^2}\right)=
\alpha_1\frac{a^2b}{c^3} +\alpha_2\frac{ab^2}{c^3} +2\alpha_3\frac{ab}{c^2}
\,,  \
\label{vaccum3}
\end{eqnarray}
and the acceleration equations (\ref{014})-(\ref{016}) reduce to
\begin{eqnarray}
&&\label{vaccum1acc}
\frac{M_1^2}{N_a^2}
\left(N_a^2\frac{k}{a^2}+2\dot{H}_a+3H_a^2-2H_a\frac{\dot{N}_a}{N_a}\right)
 =\ \ \ \ \nonumber\\
&& \ \   \frac{2\alpha_1}{3}\left(\frac{bc}{a^2}
+\frac{bN_c}{aN_a}+\frac{cN_b}{aN_a}\right)
+\frac{\alpha_2}{3}\left(\frac{b^2N_c}{a^2N_a}+2\frac{bcN_b}{a^2N_a}\right)
\nonumber\\ 
&&
+\frac{\alpha_3}{3}\left(\frac{c^2N_b}{a^2N_a}+2\frac{bcN_c}{a^2N_a}\right),
\\
\label{vaccum2acc}
&&\frac{M_2^2}{N_b^2}
\left(N_b^2\frac{k}{b^2}+2\dot{H}_b+3H_b^2-2H_b\frac{\dot{N}_b}{N_b}\right)
 =\ \ \ \ \nonumber\\
&& \ \  
\frac{\alpha_1}{3}\left(\frac{a^2N_c}{b^2N_b}+2\frac{acN_a}{b^2N_b}\right)
+\frac{2\alpha_2}{3}\left(\frac{ac}{b^2}+\frac{aN_c}{bN_b}
+\frac{cN_a}{bN_b}\right) \nonumber\\ 
&&
+\frac{\alpha_3}{3}\left(\frac{c^2N_a}{b^2N_b}+2\frac{acN_c}{b^2N_b}\right),
\\
\label{vaccum3acc}
&&\frac{M_3^2}{N_c^2}
\left(N_c^2\frac{k}{c^2}+2\dot{H}_c+3H_c^2-2H_c\frac{\dot{N}_c}{N_c}\right)
 =\ \ \ \ \nonumber\\
&& \ \   
\frac{\alpha_1}{3}\left(\frac{a^2N_b}{c^2N_c}+2\frac{abN_a}{c^2N_c}\right) 
+\frac{\alpha_2}{3}\left(\frac{b^2N_a}{c^2N_c}+2\frac{abN_b}{c^2N_c}\right)
\nonumber\\ 
&&
+\frac{2\alpha_3}{3}\left(\frac{ab}{c^2}+\frac{aN_b}{cN_c}
+\frac{bN_a}{cN_c}\right) \,.
\end{eqnarray}
In the following subsections we extract analytical solutions of the above
equations.

\subsubsection{Analytic Solutions: General Considerations}

In order to solve the cosmological equations
(\ref{vaccum3})-(\ref{vaccum3acc}) it proves more convenient to impose
specific ans\"atze. Firstly, we reparametrize the time $t$ setting
$N_a=1$. Additionally, we set
\begin{align}
N_b=\frac{\dot b}{\dot a} \quad\mbox{and}\quad N_c=\frac{\dot c}{\dot a} \,,
\label{023}
\end{align}
which directly generalize the usual assumption one makes in
bi-metric gravity in order to satisfy the Bianchi constraint
(\ref{cons_sum}). 
In particular,  substituting the functions (\ref{023}) inside the Bianchi
constraint 
(\ref{cons_sum}) leads to an automatic satisfaction, that is  
these ans\"atze are a
good starting point for our theory too. Finally, it proves convenient to
define
\begin{align}
\mathcal{B}=\frac{b}{a} \quad\mbox{and}\quad \mathcal{C}=\frac{c}{a}.
\label{BCdef}
\end{align}
Thus, the first Friedmann equations
(\ref{vaccum3}) reduce to
\begin{align}
3M_1^2\left(\frac{k}{a^2}+ H_a^2\right)&= 2\alpha_1\mathcal{B}\mathcal{C}
+\alpha_2\mathcal{B}^2\mathcal{C}+\alpha_3\mathcal{B}\mathcal{C}^2
\label{025} \,,\\
3M_2^2\left(\frac{k}{a^2}+ H_a^2\right)&=
\alpha_1\frac{\mathcal{C}}{\mathcal{B}}+2 \alpha_2\mathcal{C}
+\alpha_3\frac{\mathcal{C}^2}{\mathcal{B}} \,,\\
3M_3^2\left(\frac{k}{a^2}+ H_a^2\right)&=
\alpha_1\frac{\mathcal{B}}{\mathcal{C}}
+\alpha_2\frac{\mathcal{B}^2}{\mathcal{C}}+2 \alpha_3\mathcal{B} \,.
\label{026}
\end{align}
Subtraction of these equations eliminates the left-hand sides, and then we
can furthermore eliminate $\mathcal{B}$, obtaining a polynomial equation for
$\mathcal{C}$, which in general is of seventh order. If $\mathcal{C}\neq
M_1/M_3$, which is true in general, this equation writes as
\begin{align}
z_0 +z_1\mathcal{C} +z_2\mathcal{C}^2 +z_3\mathcal{C}^3 +z_4\mathcal{C}^4
+z_5\mathcal{C}^5 +z_6\mathcal{C}^6 +z_7\mathcal{C}^7 = 0 \,,
\label{024}
\end{align}
with the constant coefficients given by
\begin{align}
&
z_0=M_1^4\alpha_1\left(M_2^2\alpha_1^2+M_1^2\alpha_2^2\right) \,, 
\nonumber\\&
z_1=3\frac{\alpha_3}{\alpha_1}z_0\,, \qquad
z_2=-M_1^2M_3^2\alpha_1\left(4M_2^2\alpha_1^2+5M_1^2\alpha_2^2\right)\,,
\nonumber\\&
z_3= -M_1^2\alpha_3 \left(7M_1^2M_3^2\alpha_2^2
+6M_2^2M_3^2\alpha_1^2+4M_1^2M_2^2\alpha_3^2\right)\,, \nonumber\\&
z_4=M_3^2\alpha_1\left(7M_1^2M_3^2\alpha_2^2
+4M_2^2M_3^2\alpha_1^2+6M_1^2M_2^2\alpha_3^2\right)\,, \nonumber\\&
z_5=M_1^2M_3^2\alpha_3\left(5M_3^2\alpha_2^2+4M_2^2\alpha_3^2\right)\,,
\nonumber\\&
z_6=3\frac{\alpha_1}{\alpha_3}z_7\,, \qquad
z_7=-M_3^4\alpha_3\left(M_3^2\alpha_2^2+M_2^2\alpha_3^2\right)\,.
\end{align}

The existence of a real solution for $\mathcal{C}$ of the algebraic
equation (\ref{024}) is a necessary (but not sufficient) condition for the
existence of a cosmological solution. In particular, inserting  $\mathcal{C}$
into (\ref{025})-(\ref{026}) provides $\mathcal{B}$ and the scale factor
$a$. Note that the fact that $\mathcal{C}$ and $\mathcal{B}$ are constants
implies that  $c\propto b\propto a$. Moreover, the acceleration
equations (\ref{vaccum1acc})-(\ref{vaccum3acc})
will be then automatically satisfied, since the ans\"atze (\ref{023})
satisfy the Bianchi constraint (\ref{cons_sum}) which directly follows from
the field equations. Finally, we stress here that in the present
three-vierbein cosmology the seventh-order equation (\ref{024}) does not
always have a real solution, while in bi-metric gravity the corresponding
equation is of fourth order and always admits a real solution
\cite{vonStrauss:2011mq}. 

In general, the extraction of the solutions of (\ref{024}) is a hard
task. Since in this work we do not desire to rely on numerical elaboration,
in the following subsections we focus on specific simple parameter choices
which allow for analytical solutions, with however very interesting
cosmological implications. Lastly, without loss of generality, in the
following we assume the positivity of the Planck masses: $M_1>0$, $M_2>0$ and
$M_3>0$.

\subsubsection{Solutions with two $\alpha$'s being zero}
\label{twoalphaszero}

We first examine the case where two $\alpha$'s are zero. Assuming that
$\alpha_1=\alpha_2=0$, equations (\ref{025})-(\ref{024}) lead
straightforwardly to
\begin{align}
\mathcal{B}=\frac{M_1}{M_2} \,,\quad\mbox{and}\quad
\mathcal{C}=\sqrt{2}\frac{M_1}{M_3} \,.
\label{029}
\end{align}
Thus, we obtain  $c\propto b\propto a$, and the expanding
solution will be given by
\begin{eqnarray}
&&a(t)=\frac{3}{2} \frac{k M_2 M_3^2}{a_0}
\exp\left[-\sqrt{\frac{2\alpha_3M_1}{3M_2}}\frac{t}{M_3}\right]  
+\frac{a_0}{4
\alpha_3
M_1}\exp\left[\sqrt{\frac{2\alpha_3M_1}{3M_2}}\frac{t}{M_3}\right]\,,
\label{028}
\end{eqnarray}
where $a_0$ is a constant of integration. Therefore, since the first
exponential quickly becomes sub-dominant (or it is automatically zero in
the case of a flat universe), this specific triangle interaction induces  a
standard de-Sitter universe  with an effective cosmological constant
\begin{eqnarray}
\Lambda_{eff}=\frac{1}{M_3}\sqrt{\frac{2\alpha_3M_1}{3M_2}},
\end{eqnarray}
where we have to impose $\alpha_3>0$ in order
to have an expanding scale factor.


Now, due to symmetry, the above solution (\ref{028}) can be obtained in the
case where $\alpha_1=\alpha_3=0$, with the replacements
$\alpha_2\leftrightarrow\alpha_3$ and $M_2\leftrightarrow M_3$, while instead
of (\ref{029}) we will have
\begin{align}
\mathcal{B}=\sqrt{2}\frac{M_1}{M_2} \quad\mbox{and}\quad
\mathcal{C}=\frac{M_1}{M_3} \,.
\end{align}
Again we can find a de-Sitter solution when  $\alpha_2>0$.

Finally, we examine the case $\alpha_2=\alpha_3=0$, which in principle is 
 theoretically different from the previous ones since we are setting to zero
the interactions between the vierbeins whose time coordinate has not been normalized. However, the
calculations are the same
and from  (\ref{025})-(\ref{026}) we acquire
\begin{align}
\mathcal{B}=\frac{M_1}{\sqrt{2}M_2} \quad\mbox{and}\quad
\mathcal{C}=\frac{M_1}{\sqrt{2}M_3} \,,
\end{align}
that is   $c\propto b\propto a$ and thus   an
effective cosmological constant in the evolution equations
(\ref{vaccum3})-(\ref{vaccum3acc}). The general expanding solution is given
by
\begin{eqnarray}
&&a(t)=\frac{3}{2} \frac{k M_2 M_3}{a_0}
\exp\left[-\sqrt{\frac{\alpha_1}{3M_2M_3}}t\right]  +\frac{a_0}{2
\alpha_1}
\exp\left[\sqrt{\frac{\alpha_1}{3M_2M_3}}t\right]\,,
\label{030}
\end{eqnarray}
which is structurally the same as the ones obtained above, that is it
describes an   open or flat de-Sitter expansion when  $\alpha_1>0$.

\subsubsection{Bouncing Behavior}
\label{bouncingsolutions}

Interestingly, the system allows for non-singular bouncing behavior too. For
instance, although in the case of a flat universe (\ref{028}) describes an
exact de Sitter solution, considering a closed universe ($k>0$) and going
sufficiently back in time we obtain a cosmological bounce. This can be more
transparently seen if we assume for example  
$\alpha_1=\alpha_2=0$,   $k>0$ and choose
the integration constant $a_0$ to be $a_0=\sqrt{6 k M_1M_2M_3^2\al_3}$, in
which case
the solution (\ref{028}) becomes
\be \label{cosh1}
a(t)=\sqrt{\frac{3k M_2 M_3^2}{2 \al_3 M_1}}\cosh{\lp \omega t \rp}\,,
\ee
with $\omega=\Lambda_{eff}$, while the Hubble rate as a
function
of time is
\be 
H(t) = \omega \tanh {\lp \omega t \rp}\,.
\ee
This solution describes a bouncing universe which is initially  contracting
and then turns into an expanding phase. Note that in the asymptotic past and
future
the universe exhibits a  de Sitter behavior. Thus, this solution can model
also geodesically-completed inflationary cosmologies.

Similarly to the previous paragraph, for the case $\alpha_2=\alpha_3=0$ we
can also find bouncing behavior similar to (\ref{cosh1}). If we now
choose $a_0=\sqrt{3kM_2M_3\al_1}$, then (\ref{030}) can be rewritten as
\be
a(t)=\sqrt{\frac{3k M_2 M_3}{\al_1}}\cosh{\lp\sqrt{\frac{\al_1}{3M_2 M_3}}
t\rp}\,.
\ee
Moreover, note that more general classes of non-singular behaviours, in
particular asymmetric
bounces, are easy to be obtained by choosing different $a_0$. In the above
example we just presented the simplest $\cosh$-solution  in order to
illustrate the bouncing possibility\footnote{Similar hyperbolic cosine
bounce solutions have been
discovered in asymptotically free gravity 
\cite{Biswas:2005qr,Biswas:2010zk,Biswas:2012bp} and in mass-varying
\cite{Cai:2012ag} or quasi-dilaton massive gravity \cite{Gannouji:2013rwa},
however their existence did not require curvature.
}.
We mention that in the present multi-vierbein theory it is possible to
find non-singular solutions even in vacuum, whereas  in General Relativity 
to avoid the Big Bang
singularity requires one to introduce energy-condition-violating matter
sources \cite{Nojiri:2013ru,Qiu:2013eoa}. In the following we will see that, 
assuming closed universe, such vacuum solutions are generic in multi-vierbein
theories.

\subsubsection{de-Sitter Solutions}

As it is clear from the previous paragraphs, de-Sitter solutions are
particularly common in this theory. This is a general feature arising
whenever one sets $b$ and $c$ proportional to $a$. In this case all the
interacting terms in (\ref{vaccum3}) will become constants, inducing
an overall cosmological constant for all the three equations. If we
then assume spatial flatness ($k=0$) and spend our time-reparametrization
invariance setting $N_a=1$, we will find from the first equation
of (\ref{vaccum3}) the expanding
solution
\begin{align}
a(t) = a_0 \exp \left[ \frac{t}{M_1} \sqrt{\frac{2\mathcal{B}\mathcal{C}}{3}
\left(2\alpha_1+\mathcal{B}\alpha_2+\mathcal{C}\alpha_3\right)} \right] \,,
\end{align}
where we have introduced $b=\mathcal{B}\,a$ and $c=\mathcal{C}\,a$ as in
(\ref{BCdef}). This
solution corresponds to a de-Sitter universe, with the cosmological constant
depending on the interacting parameters and the two constants $\mathcal{B}$ and $\mathcal{C}$.
However, in order to complete the solution we must also satisfy the other
cosmological equations. The remaining two Friedmann equations of
  (\ref{vaccum3}) will provide solutions for the two lapse functions $N_b$
and $N_c$, which turn out to be constants depending on the free parameters
of the theory. The three acceleration equations
(\ref{vaccum1acc})-(\ref{vaccum3acc}) transform into two algebraic relations
which determine the values of $\mathcal{B}$ and $\mathcal{C}$ in terms of the
theoretical parameters. Usually these equations imply a rather involved
expression for these constants, but it considerably simplifies for some
simple cases such as, for example, the ones considered above.

\subsubsection{Solutions with $c=a\,M_1/M_3$}

In order to extract equation (\ref{024}) we assumed that
$\mathcal{C}\neq M_1/M_3$. If this is not the case, that is if 
$\mathcal{C}=
M_1/M_3$, then equations (\ref{025})-(\ref{026}) are satisfied only if
$\alpha_1 M_3=\alpha_3M_1$, resulting to a cubic equation for $\mathcal{B}$:
\begin{eqnarray}
-2\frac{\alpha_1}{M_2^2} +-2\frac{\alpha_2}{M_2^2}\mathcal{B}
+3\frac{\alpha_1}{M_1^2}\mathcal{B}^2 +\frac{\alpha_2}{M_1^2}\mathcal{B}^3=0.
\label{027}
\end{eqnarray}
If one of $\alpha_1$ or $\alpha_3$ is zero, then the other will be too, and
thus we result in the case described in paragraph \ref{twoalphaszero}.
However, if  only $\alpha_2=0$ then equation (\ref{027}) has the unique
positive solution
\begin{align}
\mathcal{B}=\sqrt{\frac{2}{3}}\frac{M_1}{M_2} \,,
\end{align}
 that is $c\propto b\propto a$ which implies an effective
cosmological constant in the Friedmann equations  (\ref{vaccum3}). The
expanding solution will then be
\begin{eqnarray}
&&a(t)=\frac{3}{2} \frac{k M_2 M_3}{a_0}
\exp\left[-\left(\frac{2}{3}\right)^{1/4}
\sqrt{\frac{\alpha_1}{M_2M_3}}t\right] 
+\frac{a_0}{2 \sqrt{6}
\alpha_1}\exp\left[\left(\frac{2}{3}\right)^{1/4}
\sqrt{\frac{\alpha_1}{M_2M_3}}t\right],\ \ \
\label{032}
\end{eqnarray}
which becomes a pure de-Sitter universe for $k=0$ and $\alpha_1>0$, or a
late-time de Sitter universe for  $k>0$ and and $\alpha_1>0$. Additionally,
since the equations are symmetric in the vierbein exchange $2\leftrightarrow 3$, we
obtain another solution similar to (\ref{032}), with
  the substitutions $\alpha_2\leftrightarrow\alpha_3$ and
$M_2\leftrightarrow M_3$.

Finally, note that in this case we also find non-singular bouncing
evolutions. For example the choice $a_0=\sqrt{3k\sqrt{6}M_2M_3\al_1}$
corresponds to
\be
a(t)=\sqrt{\frac{\sqrt{6} kM_2M_3}{2\al_1}}\cosh{\left[
\lp\frac{2}{3}\rp^\frac{1}{4}\sqrt{\frac{\al_1}{M_2 M_3}}t\right]}\,,
\ee
which exhibits a bounce before transiting to an expanding de Sitter universe.

We close this paragraph mentioning that all the above vacuum  solutions
fulfill the Friedmann equation with a constant source. Therefore, all of
them correspond to the same family, that is to the de Sitter one (although
under specific conditions they can exhibit a bouncing behavior before
entering the pure de Sitter regime).

\subsection{Matter solutions}
\label{Sec:matterSol3}

In order to obtain a late-time description of the universe, we must take
into account the matter sector. According to the discussion of
section \ref{mattercoupling} the choice of the matter coupling will yield
constraints on the field equations.  In this section we will consider two
cases in particular and we will find simple analytical solutions to the field
equations.

\subsubsection{Matter coupled to one (physical) vierbein}

The first case we consider is when the matter sector  is coupled  to  one
tetrad only, which thus turns out to be the physical vierbein. 
This is the simplest and usual case, since it satisfies the equivalence
principle. We assume that
matter can be described by a perfect fluid with $\rho_m$ and $p_m$ its energy
density and pressure respectively.  The coupled/physical tetrad will be the
one with the scale factor $a$, that is we set $c_1=1$ and $c_2=c_3=0$
in action (\ref{actionfull}).
Focusing on the $\alpha$-terms as before, that is to the three-vertex
interaction,  
 the Friedmann equations for the physical vierbein now read as  
\begin{equation}
 3M_1^2\left(\frac{k}{a^2}+\frac{H_a^2}{N_a^2}\right)=
\rho_m+2\alpha_1\frac{b c}{a^2} +\alpha_2\frac{b^2 c}{a^3}+ \alpha_3\frac{b
c^2}{a^3},
\end{equation}
\begin{eqnarray}
 &&\frac{M_1^2}{N_a^2}
\left(N_a^2\frac{k}{a^2}+2\dot{H}_a+3H_a^2-2H_a\frac{\dot{N}_a}{N_a}\right)
= -p_m+\frac{2\alpha_1}{3}\left(\frac{bc}{a^2}
+\frac{bN_c}{aN_a}+\frac{cN_b}{aN_a}\right)
\nonumber\\ 
&& \ \ \ \ \ \ \ \ \ \ \ \ \ \ \ \ \ \ \ \ \ \ \  \ \ \ \ \ \ \ \ \ \ \ \ \ \
\ 
+\frac{\alpha_2}{3}\left(\frac{b^2N_c}{a^2N_a}+2\frac{bcN_b}{a^2N_a}\right)
+\frac{\alpha_3}{3}\left(\frac{c^2N_b}{a^2N_a}+2\frac{bcN_c}{a^2N_a}\right)
, 
\end{eqnarray}
while the corresponding equations for the other two viebeins remain the same
as in the second and third equations of (\ref{vaccum3}) and those in 
(\ref{vaccum2acc}),(\ref{vaccum3acc}).

According to the considerations we made in Sec.~\ref{mattercoupling}, due to
the diffeomorphism invariance of the action, the matter energy-momentum
tensor must be covariantly conserved. Having only one matter sector in the
present case, this implies that
\begin{align}
\dot\rho_m + 3 H_a (\rho_m+p_m)=0 \,.
\end{align}
If we now assume the standard linear equation of state $p_m=w\rho_m$, we inevitably obtain
\begin{align}
\rho_m \propto a^{-3(1+w)} \,,
\end{align}
meaning that the average matter in the universe decays as it does in General
Relativity.

In order to find a simple analytical solution we set $N_a=1$
and we impose
\begin{align}
N_b=\frac{\dot b}{\dot a} \quad\mbox{and}\quad N_c=\frac{\dot c}{\dot a} \,,
\end{align}
which in turn  guarantees that the Bianchi constraint (\ref{cons_sum}) is
satisfied. Due to the matter coupling the field equations are much harder to
be solved comparing to the vacuum case. To
further simplify the problem we impose $b=c$ and $M_2=M_3$, which actually
reduce the model to an effective bi-vierbein theory. However, with these
simplifications we manage to find an interesting solution. In particular,
assuming spacial flatness
($k=0$) and choosing $\alpha_2=\alpha_3=0$, which leads to one independent
parameter, namely $\alpha_1$, then the (non-physical) scale factors $b$ and
$c$ are related to $a$ by
\begin{align}
b = c = \frac{M_1}{\sqrt{2}M_2} a \sqrt{1-B_0\,a^{-3(w+1)}} \,,
\end{align}
with $B_0$ a constant of integration. If such a constant is set to zero the
solution reduces the the simplest case $b=c\propto a$. The physical scale
factor $a$ has the interesting solution
\begin{align}
a\propto \exp\left(\frac{t\sqrt{\alpha_1}}{M_2\sqrt{3}}\right) \,,
\end{align}
which again correspond to a de-Sitter expansion provided $\alpha_1>0$. Note
that this very specific solution allows for a de-Sitter expansion in presence
of non negligible matter. Such solutions can be relevant in characterizing
the observed transition from a matter to a dark-energy dominated era.

Due to the complexity of the field equations, finding analytical solutions
is difficult. Therefore, in order to extract more realistic evolutions, with the
correct quantitative behavior of matter and dark-energy epochs, one needs to
resort to numerical elaboration. Since in the present work we desire to
retain the investigations analytical, this analysis is left for a
future project.

\subsubsection{Matter coupled to all vierbeins}

As we discussed in the beginning of section \ref{mattercoupling}, in
principle one can go beyond the simple one-metric-matter coupling of the
previous paragraph, and consider   couplings to more vierbeins
simultaneously. Since such couplings violate the equivalence principle they
are potentially strongly
constrained by experiments\footnote{However due to the Vainshtein screening mechanism it is not clear how strong 
the constraints will turn out.}, however apart from this experimental requirement,
at the theoretical level they appear to be allowed, and thus it would be interesting
to consider them too as a first step towards understanding their
implications. Therefore, in the present example we assume that the matter
sector couples to the three-vierbein gravity in a completely symmetric way,
that is we consider the case where the matter action is composed by
three equal Lagrangians coupling to different tetrads. The total action is
given by (\ref{actionfull}), with all $c_i$ being non-zero.
Again, we restrict our analysis to the case where only the
$\alpha$-terms are nonvanishing.

Assuming that the three matter sectors are given by a perfect fluid
energy-momentum tensor, the first Friedmann equations become  
\begin{eqnarray}
&&3M_1^2\left(\frac{k}{a^2}+\frac{H_a^2}{N_a^2}\right)=c_1\rho_1+ 2\alpha_1\frac{b
c}{a^2} +\alpha_2\frac{b^2 c}{a^3}+ \alpha_3\frac{b c^2}{a^3}  
,\nonumber\\
&&3M_2^2\left(\frac{k}{b^2}+\frac{H_b^2}{N_b^2}\right)=c_2\rho_2+
\alpha_1\frac{a^2c}{b^3} +2\alpha_2\frac{ac}{b^2} +\alpha_3\frac{ac^2}{b^3}
,\nonumber\\
&&3M_3^2\left(\frac{k}{c^2}+\frac{H_c^2}{N_c^2}\right)=c_3\rho_3+
\alpha_1\frac{a^2b}{c^3} +\alpha_2\frac{ab^2}{c^3} +2\alpha_3\frac{ab}{c^2}
\,, \nonumber\\ 
\end{eqnarray}
and the acceleration equations generalize to
\begin{eqnarray}
&&
\frac{M_1^2}{N_a^2}
\left(N_a^2\frac{k}{a^2}+2\dot{H}_a+3H_a^2-2H_a\frac{\dot{N}_a}{N_a}\right)
 =-c_1p_1\ \ \ \ \nonumber\\
&& \ \   +\frac{2\alpha_1}{3}\left(\frac{bc}{a^2}
+\frac{bN_c}{aN_a}+\frac{cN_b}{aN_a}\right)
+\frac{\alpha_2}{3}\left(\frac{b^2N_c}{a^2N_a}+2\frac{bcN_b}{a^2N_a}\right)
\nonumber\\ 
&&
+\frac{\alpha_3}{3}\left(\frac{c^2N_b}{a^2N_a}+2\frac{bcN_c}{a^2N_a}\right),
\\
&&\frac{M_2^2}{N_b^2}
\left(N_b^2\frac{k}{b^2}+2\dot{H}_b+3H_b^2-2H_b\frac{\dot{N}_b}{N_b}\right)
 =-c_2p_2\ \ \ \ \nonumber\\
&& \ \  
+\frac{\alpha_1}{3}\left(\frac{a^2N_c}{b^2N_b}+2\frac{acN_a}{b^2N_b}\right)
+\frac{2\alpha_2}{3}\left(\frac{ac}{b^2}+\frac{aN_c}{bN_b}
+\frac{cN_a}{bN_b}\right) \nonumber\\ 
&&
+\frac{\alpha_3}{3}\left(\frac{c^2N_a}{b^2N_b}+2\frac{acN_c}{b^2N_b}\right),
\\
&&\frac{M_3^2}{N_c^2}
\left(N_c^2\frac{k}{c^2}+2\dot{H}_c+3H_c^2-2H_c\frac{\dot{N}_c}{N_c}\right)
 =-c_3p_3\ \ \ \ \nonumber\\
&& \ \   
+\frac{\alpha_1}{3}\left(\frac{a^2N_b}{c^2N_c}+2\frac{abN_a}{c^2N_c}\right) 
+\frac{\alpha_2}{3}\left(\frac{b^2N_a}{c^2N_c}+2\frac{abN_b}{c^2N_c}\right)
\nonumber\\ 
&&
+\frac{2\alpha_3}{3}\left(\frac{ab}{c^2}+\frac{aN_b}{cN_c}
+\frac{bN_a}{cN_c}\right) \,,
\end{eqnarray}
where $\rho_i$ and $p_i$ are the energy density and pressure of  the fluid
coupled to the $i$th tetrad.

These equations are even more difficult to be handled than the ones arising
in the previous matter case. Since a numerical analysis is beyond the purpose
of the present paper, we will limit our discussion to an almost trivial
analytical solution. This is achieved setting $c\propto b\propto a$ and the
three matter sector to coincide, that is $\rho_1=\rho_2=\rho_3=\rho$ and
$c_1=c_2=c_3=1$. The model reduces   to an effective single-tetrad theory
if we additionally assume $M_1=M_2=M_3$ and, after imposing the equation of
state $p_i=w\rho_i$, a simple solution can be easily found in the $k=0$ case.
For   $\alpha_2=\alpha_3=0$  and $\alpha_1<0$
this can be written as
\begin{align}
a \propto \cos\left[\frac{\sqrt{-3\,\alpha_1}}{2M_1}(w+1)(t-t_0)\right]^{\frac{2}{3(w+1)}} \,,
\end{align}
where $t_0$ is a constant of integration. The solution for the matter
evolution is instead given by
\begin{align}
\rho = -\alpha_1
\sec\left[\frac{\sqrt{-3\,\alpha_1}}{2M_1}(w+1)(t-t_0)\right]^{2} \,,
\label{034}
\end{align}
which describes a universe with a Big Bang and a Big Crunch symmetrically
situated in the past and future of $t=t_0$. Lastly, note that (\ref{034})
implies that the matter energy density becomes infinite at the
two singularities.



\section{Four-Vierbein Cosmology}
\label{fourvierbein}

\begin{figure}[ht]
\subfloat[][Full interacting
theory]{\includegraphics[width=0.45\columnwidth]{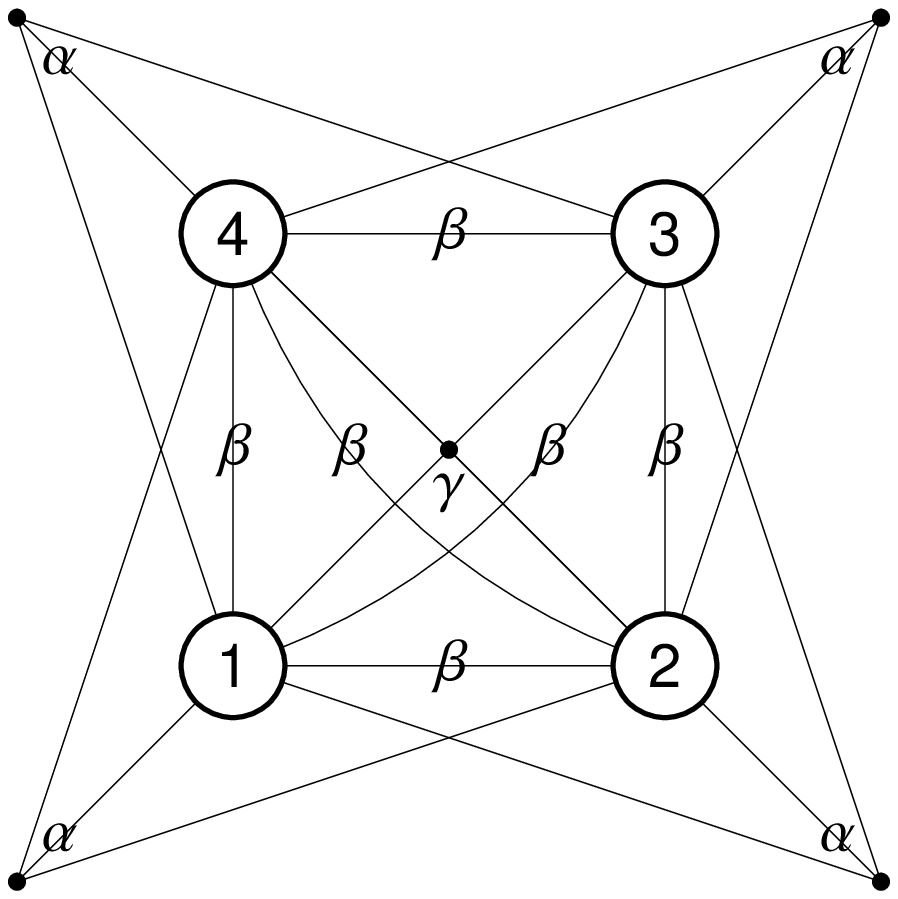}} \qquad
\subfloat[][Four-vertex interacting
theory]{\includegraphics[width=0.45\columnwidth]{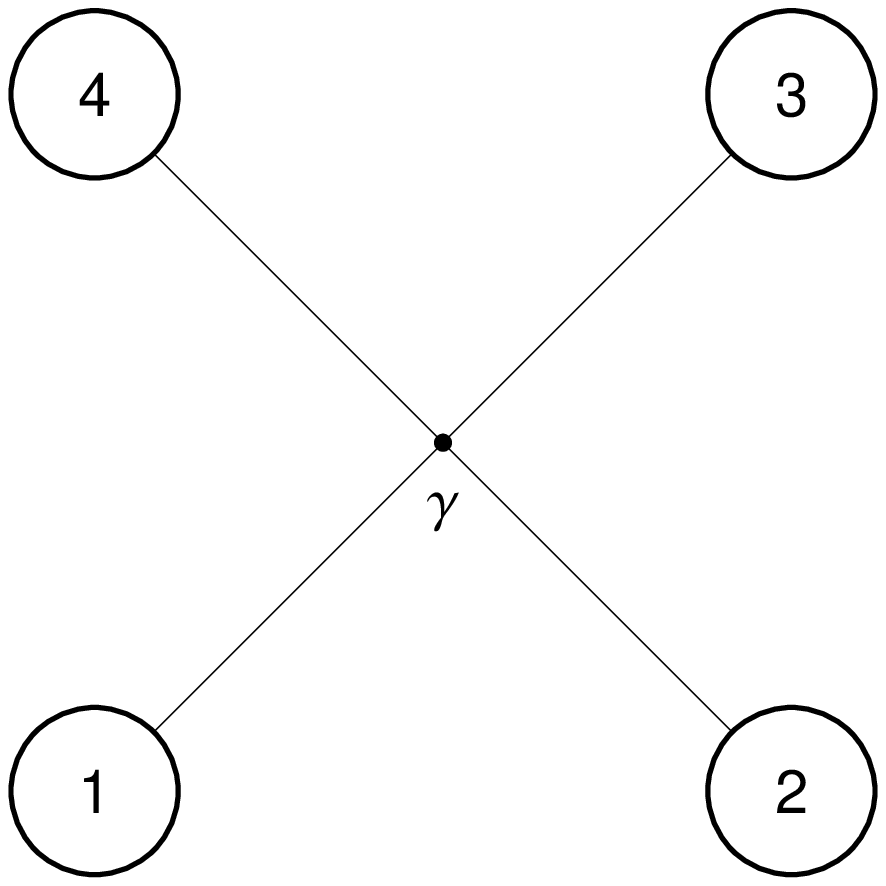}}
\caption{
Graphic representation of the four-vierbein theory in four
 dimensions. Bi-tetrad interactions are labeled by $\beta$-edges,
three-tetrad interactions  by $\alpha$-three-vertices and four-tetrad
interactions by the $\gamma$-four-vertex. The $\Lambda$-terms (cosmological
constants)  
correspond to self-interacting/closed lines and have not be drawn.
}
\label{fig:2}
\end{figure}

In this section we will consider four interacting spin-2 fields in four
dimensions. The number of the coupling terms is now 35: there are 
4 cosmological constant terms
($\Lambda_i$), 18 terms coupling pairs of metrics ($\beta_i$), 12
three-vertex
terms ($\alpha_i$) and just one term  mixing all the four metrics
($\gamma$).
The graphical visualization of the full theory is given in
Fig.~\ref{fig:2}~(a). However, since we are interested in examining the
case where there is a maximal interaction between the vierbeins, in this
section we  focus on the $\gamma$-term, setting all the other
parameters to zero, that is we study the  model depicted in
Fig.~\ref{fig:2}~(b). Clearly, the investigation of the full theory is
straightforward.

Similarly to the previous section, in order to proceed to the cosmological
applications we consider a diagonal  FRW ansatz for  all the four
tetrads (again this means that for the following solutions the  Deser-van
Nieuwenhuizen condition is satisfied):
\begin{align} 
e_{(1)}{}_\mu^a=\mbox{diag}\left(N_a(t), \frac{a(t)}{\sqrt{1-k\,r^2}}, a(t)\,
r, a(t)\,r\,\sin\theta\right)  \,,\nonumber\\
e_{(2)}{}_\mu^a=\mbox{diag}\left(N_b(t), \frac{b(t)}{\sqrt{1-k\,r^2}}, b(t)\,
r, b(t)\,r\,\sin\theta\right) \,,\nonumber\\
e_{(3)}{}_\mu^a=\mbox{diag}\left(N_c(t), \frac{c(t)}{\sqrt{1-k\,r^2}}, c(t)\,
r, c(t)\,r\,\sin\theta\right)  \,,\nonumber\\
e_{(4)}{}_\mu^a=\mbox{diag}\left(N_d(t), \frac{d(t)}{\sqrt{1-k\,r^2}}, d(t)\,
r, d(t)\,r\,\sin\theta\right) \,,
\end{align}
corresponding to the
metrics
\begin{align}
ds^2_1 = -N_a(t)^2\,dt^2 + a(t)^2 \left(\frac{dr^2}{1-k r^2}+r^2
d\Omega^2\right)
 \,,\nonumber\\
ds^2_2 = -N_b(t)^2\,dt^2 + b(t)^2 \left(\frac{dr^2}{1-k r^2}+r^2
d\Omega^2\right)
\,,\nonumber\\
ds^2_3 = -N_c(t)^2\,dt^2 + c(t)^2 \left(\frac{dr^2}{1-k r^2}+r^2
d\Omega^2\right)
 \,,\nonumber\\
ds^2_4 = -N_d(t)^2\,dt^2 + d(t)^2 \left(\frac{dr^2}{1-k r^2}+r^2
d\Omega^2\right)
\,.
\end{align}
Action variation provides the four first Friedmann equations
\begin{align}
\frac{\gamma}{M_1^2}\frac{b c d}{a^3}=\frac{k}{a^2}+\frac{H_a^2}{N_a^2}
\,,\quad
\frac{\gamma}{M_2^2}\frac{a c d}{b^3}=\frac{k}{b^2}+\frac{H_b^2}{N_b^2}
 \,,\nonumber\\
\frac{\gamma}{M_3^2}\frac{a b d}{c^3}=\frac{k}{c^2}+\frac{H_c^2}{N_c^2}
\,,\quad
\frac{\gamma}{M_4^2}\frac{a b c}{d^3}=\frac{k}{d^2}+\frac{H_d^2}{N_d^2} \,,
\end{align}
and the four acceleration equations
\begin{eqnarray}
&& 2 \dot{H}_a+3 H_a^2-2 H_a\frac{\dot{N}_a}{N_a}+\frac{k}{a^2}N_a^2
=\nonumber\\
&&\ \ \ \ \ \ \ \ \ \ \frac{\gamma}{M_1^2}\left(\frac{b c}{a^2} N_a
N_d+\frac{b d}{a^2} N_a
N_c+\frac{c d}{a^2} N_a N_b\right) ,\nonumber\\
&&2 \dot{H}_b+3 H_b^2-2 H_b\frac{\dot{N}_b}{N_b}+\frac{k}{b^2}N_b^2
=\nonumber\\
&&\ \ \ \ \ \ \ \ \ \ 
\frac{\gamma}{M_2^2}\left(\frac{a c}{b^2} N_b N_d+\frac{a d}{b^2} N_b
N_c+\frac{c d}{b^2} N_a N_b\right) ,\nonumber\\
&&2 \dot{H}_c+3 H_c^2-2 H_c\frac{\dot{N}_c}{N_c}+\frac{k}{c^2}N_c^2
=\nonumber\\
&&\ \ \ \ \ \ \ \ \ \ 
\frac{\gamma}{M_3^2}\left(\frac{a b}{c^2} N_c N_d+\frac{a d}{c^2} N_b
N_c+\frac{b d}{c^2} N_a N_c\right) ,\nonumber\\
&&2 \dot{H}_d+3 H_d^2-2
H_d\frac{\dot{N}_d}{N_d}+\frac{k}{d^2}N_d^2=\nonumber\\
&&\ \ \ \ \ \ \ \ \ 
\frac{\gamma}{M_4^2}\left(\frac{a b}{d^2} N_c N_d+\frac{a c}{d^2} N_b
N_d+\frac{b c}{d^2} N_a N_d\right) ,
\end{eqnarray}
where the parameter $\gamma$ is related to the   completely symmetric
tensor of
coefficients $\chi^{ijkl}$ of equation (\ref{003})  through  
\begin{align}
\gamma = 12\,\chi^{1234} \,.
\end{align}

In what follows we will consider  simple analytical solutions of
the four-tetrad theory in vacuum. For simplicity we will not extract matter
solutions, having in mind that the analysis of
Sec.~\ref{Sec:matterSol3} can be generalized here too.

\subsection{Vacuum solutions}

In this case we can apply the procedure of subsection \ref{vacuumsolthree} of
the three-viebein theory. Doing so we result in one solution (since we have
just
one interacting term), with $d\propto c\propto b\propto a$ and $N_b=b$,
$N_c=c$, $N_d=d$. In order
to satisfy the field equations the constants of proportionality must be
\begin{align}
b=\frac{M_1}{M_2}\,a \,,\qquad c=\frac{M_1}{M_3}\,a \,,\qquad
d=\frac{M_1}{M_3}\,a \,.
\end{align}
Then setting $N_a=1$ the  solution for $a$ is
\begin{eqnarray}
a(t)=\frac{k M_2 M_3 M_4}{a_0} \exp\left[-\sqrt{\frac{\gamma
M_1}{M_2M_3M_4}}t\right] 
+\frac{a_0}{4 \gamma
M_1}\exp\left[\sqrt{\frac{\gamma
M_1}{M_2M_3M_4}}t\right]\,.
\label{033}
\end{eqnarray}
This reduces to a de-Sitter expansion if we consider $k=0$, with $\gamma>0$.

All the considerations made in Sec.~\ref{model} are generally valid
also for this solution.
In particular, non-singular bouncing evolution can be realized in vacuum
by choosing $a_0=\sqrt{4\gamma k M_1M_2M_3M_4}$, leading to
\be
a(t)= \sqrt{\frac{ k M_2M_3 M_4}{\gamma M_1}}\cosh{\lp \sqrt{\frac{\gamma
M_1}{M_2M_3M_4}}t\rp}\,,
\ee
which exhibits a bounce before transiting to an expanding de Sitter universe.
Thus, the existence of such behavior seems to be a generic feature of
multi-tetrad theories.

\section{Discussion and Conclusions}
\label{Conclusions}

In this work we investigated the cosmology of interacting spin-2 particles.
We formulated the full theory in terms of vierbeins, but without imposing 
the Deser-van Nieuwenhuizen  constraint or any similar restriction, which in
the general case is not a result of the field equations themselves. Since the
 imposition of such a restriction assures the
equivalence of metric and vielbein formulations of bi-metric theories
\cite{Hassan:2012wt}, its absence implies that the resulting multi-vierbein
theory is different and much richer than the corresponding multi-metric
theory, of which is not even known whether it exists or not. Secondly, since the
ghost-freedom for more than two tetrads can be proven only in the vierbein formulation
\cite{Hinterbichler:2012cn}, while in the metric description such a
general proof does not exist for the moment, even if one finds a way to construct the
multi-metric correspondent of the above general multi-vierbein theory, it is
not guaranteed that it will be ghost free. The un-restricted multi-vielbein
formulation seems to describe a much wider class of theories, which can be
used to characterize interacting gravitational sectors. Finally, in order to
study the cosmological applications,  we introduced the coupling to the matter
sector in a self-consistent way.

We studied the cases of three or four interacting vierbeins, focusing on
the novel multi-interacting terms that do not have a known multi-metric
formulation, setting all the other interacting and non-interacting terms to
zero. Clearly, one can study the full interacting theory, or theories with
more vierbeins, straightforwardly. 

In the case of vacuum solutions  we found many de-Sitter expansions, where
the effective cosmological constant arises solely from the combination of
the multi-interacting terms. Such solutions have a great physical impact
since they can describe the inflationary era. In the case where matter is
present  we found accelerating solutions, which can describe the dark-energy
epoch. Additionally, for particular parameter choices   we found bouncing  
behavior. 

Finally, we mention that the great complexity that arise in a
theory with three or more tetrads does not allow for an analytical
treatment of more convoluted cosmological solutions. In order to proceed
beyond the extraction of simple and basic analytical solutions one needs to 
perform numerical elaboration, and indeed in this case he can obtain a richer
cosmological behavior, closer to the detailed cosmological history.
However, such a detailed numerical investigation lies beyond the aim of
the present work, which is to define the cosmology of such
theories  and   to show that at least simple and basic analytical solutions
can be constructed.

The above analysis shows that the un-restricted multi-viebein cosmology is
richer and includes novel features comparing to bi-metric gravity. Clearly,
before accepting such constructions as candidates for the description of
nature, many additional investigations should be performed, amongst others
the use of observational data in order to constrain the parameters of the
theory, a detailed dynamical analysis that could reveal its asymptotic
features and the systematic study of the perturbations. These
investigations, although necessary, lie beyond the scope of the present work
and are left for future investigations.


\begin{acknowledgments}
The authors wish to thank S.~Deser, S.~Capozziello, F.~Hassan, R.~Rosen and S.~Speziale for useful discussions. 
The research
of ENS is implemented
within the framework of the Action ``Supporting Postdoctoral Researchers''
of the Operational Program ``Education and Lifelong Learning'' (Actions
Beneficiary: General Secretariat for Research and Technology), and is
co-financed by the European Social Fund (ESF) and the Greek State.
TK is supported by the Norwegian Research Council of Norway.
\end{acknowledgments}
\appendix

\section{Derivation of the field equations}
\label{app:A}

Consider action (\ref{001}) which we recall for the sake of simplicity
\begin{align}
S = \int \sum_i \mathcal{L}^{(i)}_{\rm EH} + U \,,
\label{002}
\end{align}
with
\begin{align}
U=\epsilon_{abcd} \sum_{i,j,k,l} \tilde\chi^{ijkl}\, {\bf e}_{(i)}^a \wedge
{\bf e}_{(j)}^b \wedge {\bf e}_{(k)}^c \wedge {\bf e}_{(l)}^d \,.
\end{align}
In order to extract the field equations for the $i$th tetrad we must vary the
action with
respect to $e_{(i)}$. The first term in (\ref{002}) will produce the usual
Einstein tensor and its variation will not be considered here. The variation
of $U$ is
\begin{align}
\delta_{(i)} U = \epsilon_{abcd} \sum_{j,k,l} \chi^{ijkl}\, \delta{\bf
e}_{(i)}^a \wedge {\bf e}_{(j)}^b \wedge {\bf e}_{(k)}^c \wedge {\bf
e}_{(l)}^d \,,
\end{align}
where $\chi^{ijkl}=\mathcal{P}(i)\tilde\chi^{ijkl}$, with $\mathcal{P}(i)$
denoting the number of times the $i$th index appears in $\tilde\chi^{ijkl}$.
This takes into account the number of times $e_{(i)}$ appears inside one
term.
In other words $\mathcal{P}(i)$ is 1 for ${\bf e}_{(i)}^a \wedge {\bf
e}_{(j)}^b \wedge {\bf e}_{(k)}^c \wedge {\bf e}_{(l)}^d$, 2 for ${\bf
e}_{(i)}^a \wedge {\bf e}_{(i)}^b \wedge {\bf e}_{(j)}^c \wedge {\bf
e}_{(k)}^d$, 3 for ${\bf e}_{(i)}^a \wedge {\bf e}_{(i)}^b \wedge {\bf
e}_{(i)}^c \wedge {\bf e}_{(j)}^d$ and 4 for ${\bf e}_{(i)}^a \wedge {\bf
e}_{(i)}^b \wedge {\bf e}_{(i)}^c \wedge {\bf e}_{(i)}^d$, with $j,k,l$
taking all possible values from 1 to $N$.

At this point Eq.~(\ref{003}) follows immediately. To rewrite it as
Eq.~(\ref{004}) note that
\begin{align}
\delta_{(i)} U &= \epsilon_{abcd} \sum_{j,k,l} \chi^{ijkl}\, \delta{\bf
e}_{(i)}^a \wedge {\bf e}_{(j)}^b \wedge {\bf e}_{(k)}^c \wedge {\bf
e}_{(l)}^d \nonumber\\
&= \epsilon_{abcd} \sum_{j,k,l} \chi^{ijkl}\, \delta e_{(i)\mu}^a
e_{(j)\nu}^b e_{(k)\sigma}^c e_{(l)\rho}^d \epsilon^{\mu\nu\sigma\rho} d^4x
\nonumber\\
&=\delta e_{(i)\mu}^a\,e_{(i)e}^\mu\,(\det e_{(i)}d^4x)\,
\epsilon_{abcd}\,\epsilon^{efgh}   \sum_{j,k,l} \chi^{ijkl}\,
(e_{(i)f}^\nu
e_{(j)\nu}^b) (e_{(i)g}^\sigma e_{(k)\sigma}^c) (e_{(i)h}^\rho e_{(l)\rho}^d)
\nonumber\\
&=\delta e_{(i)\mu}^a\,e_{(i)e}^\mu\,(\det e_{(i)}d^4x)\,
\epsilon_{abcd}\,\epsilon^{efgh}   \sum_{j,k,l} \chi^{ijkl}\,
X_{(ij)}{}_f{}^b
X_{(ik)}{}_g{}^c X_{(il)}{}_h{}^d \,.
\end{align}
Adding to this the usual variation of the Einstein-Hilbert Lagrangian and
then contracting with $e_{(i)\nu}^a$, will eventually provide
Eq.~(\ref{004}).

\end{document}